\documentclass[12pt,draftclsnofoot,journal,onecolumn]{IEEEtran}

\IEEEoverridecommandlockouts

\usepackage{graphicx}
\usepackage{cite}
\usepackage{url}
\usepackage[cmex10]{amsmath}
\usepackage{amssymb}
\usepackage{algorithm}
\usepackage[noend]{algorithmic}
\usepackage{cases}
\usepackage[caption=false,font=footnotesize]{subfig}
\usepackage{color}

\DeclareMathOperator{\cov}{cov}
\DeclareMathOperator{\diag}{diag}

\begin{document}

\title{Multiple Antenna Cyclostationary Spectrum Sensing Based on the Cyclic Correlation Significance Test}
\author{Paulo~Urriza, Eric~Rebeiz, Danijela~Cabric%
\thanks{The authors are with the Electrical Engineering Department, University of California, Los Angeles, CA, 90095, USA (email: \{pmurriza, rebeiz, danijela\}@ee.ucla.edu).}
\thanks{Part of this work has been accepted to the proceedings of the Global Communications Conference (GLOBECOM), Dec. 3--7, 2012, Anaheim, CA, USA \cite{Urriza2012}.}
\thanks{This work has been submitted to the IEEE for possible publication. Copyright may be transferred without notice, after which this version may no longer be accessible.}
}
\maketitle

%%%%%%%%%%%%%%%%%%%%%%%%%%%%%%%%%%%%%%%%%%%%%%%%%%%%%%%%%%%%%%%%%%%%%%%%%%%%%%%
%%%%%%%%%%%%%%%%%%%%%%%%%%%%%%%%%%%%%%%%%%%%%%%%%%%%%%%%%%%%%%%%%%%%%%%%%%%%%%%

\begin{abstract}
In this paper, we propose and analyze a spectrum sensing method based on cyclostationarity specifically targeted for receivers with multiple antennas. This detection method is used for determining the presence or absence of primary users in cognitive radio networks based on the eigenvalues of the cyclic covariance matrix of received signals. In particular, the cyclic correlation significance test is used to detect a specific signal-of-interest by exploiting knowledge of its cyclic frequencies. Analytical expressions for the probability of detection and probability of false-alarm under both spatially uncorrelated or spatially correlated noise are derived and verified by simulation. The detection performance in a Rayleigh flat-fading environment is found and verified through simulations. One of the advantages of the proposed method is that the detection threshold is shown to be independent of both the number of samples and the noise covariance, effectively eliminating the dependence on accurate noise estimation. The proposed method is also shown to provide higher detection probability and better robustness to noise uncertainty than existing multiple-antenna cyclostationary-based spectrum sensing algorithms under both AWGN as well as a quasi-static Rayleigh fading channel.
\end{abstract}

\IEEEpeerreviewmaketitle

%\begin{keywords}
%\end{keywords}

%%%%%%%%%%%%%%%%%%%%%%%%%%%%%%%%%%%%%%%%%%%%%%%%%%%%%%%%%%%%%%%%%%%%%%%%%%%%%%%
%%%%%%%%%%%%%%%%%%%%%%%%%%%%%%%%%%%%%%%%%%%%%%%%%%%%%%%%%%%%%%%%%%%%%%%%%%%%%%%

\section{Introduction}
\label{sec:Introduction}

%Spectrum sensing importance in cognitive radio
Spectrum sensing is a key step in effectively realizing cognitive radio networks (CRN). In the CR access paradigm, secondary users (SU) in a CRN are allowed to access spectrum reserved for use by licensed or primary users (PU) given that 1) those resources are either currently unoccupied or 2) interference to the primary network is kept under an acceptable level \cite{Haykin2005}. The main goal of spectrum sensing is to accurately and efficiently detect the presence or absence of a PU in a given band, usually under the constraint of a low signal-to-noise ratio (SNR).

%General categories of spectrum sensing focusing on cyclostationary-based
Several spectrum sensing methods have been proposed in the literature \cite{Yucek2009}. In general, these methods can be categorized as being based on either energy detection, cyclic correlation (cyclostationarity), or matched filtering. Energy detection requires the least prior knowledge about the signal, while matched filtering requires the most. Cyclic correlation-based techniques lie in between, requiring either prior knowledge or accurate estimation of the cyclic frequencies present in the PU transmission signal. Although energy detection offers the lowest computational complexity and is the optimal blind detector in the presence of i.i.d. noise, its performance relies on accurate knowledge of noise power due to the SNR wall phenomenon \cite{Tandra2008}. The detection performance of energy detection also degrades in a temporally correlated noise environment.

In some scenarios, such as very low SNR regime or when signal selectivity is important, cyclic correlation-based methods offer several advantages over other spectrum sensing approaches. Unlike energy detection, they do not suffer from the SNR wall issue. These methods are also resilient to temporally correlated noise and enable signal-selective spectrum sensing where the presence of signals-of-interest (SOI) can be detected based on their unique cyclic features due to their modulation type, symbol rate, and carrier frequency \cite{Gardner1987a}. 

%Multiple antenna advantages for cyclostationarity and algorithms proposed
One issue encountered with all spectrum sensing methods is the effect of fading in the channel between the PU and SU. There is a decrease in the probability of detection whenever the channel is in a deep fade. This can be alleviated by exploiting spatial diversity either through the use of cooperative spectrum sensing \cite{Quan2008} or, if available, the use of multiple antennas. As a result, spectrum sensing algorithms exploiting multiple antennas have received considerable interest \cite{Taherpour2010,Tugnait2012}. 

%In this paper (key advantages)
Algorithms that leverage the cyclostationarity property have been applied in the past for multiple antenna receivers. In \cite{Sadeghi2008}, the sum of the spectral correlation for each antenna was proposed. Such methods are considered \textit{post-combining} techniques since knowledge of the channel state information (CSI) is not exploited. On the other hand, \textit{pre-combining} techniques which utilize an estimate of the CSI to varying degrees have been shown to have better performance in a random channel. A method based on equal gain combining (EGC) was investigated in \cite{Chen2008} which uses phase offset estimates to align the raw samples from each antenna. The aligned signals are then summed before finding the spectral correlation. Finally, a blind maximal ratio combining (MRC) scheme was evaluated in \cite{Jitvanichphaibool2010a} which utilized the singular value decomposition (SVD) to find an estimate of the CSI and applied MRC on the raw samples.

In this paper we propose a spectrum sensing algorithm based on the cyclic correlation significance test (CCST) designed for use in a multiple antenna system which we refer to as Eigenvalue-Based Cyclostationary Spectrum Sensing or EV-CSS. The CCST was used in \cite{Schell1990a} to perform cyclostationary source enumeration using an information-theoretic criterion. However, the use of CCST in the context of multiple-antenna cyclostationary spectrum sensing has not been investigated in prior work. The performance of this method in fading channels has also not been evaluated nor compared to other spectrum sensing schemes that exploit cyclostationarity. In this paper, we derive the analytical performance of this detection method in both AWGN and flat-fading channels. These expressions are then verified through simulations. The results also enable us to investigate sensing performance in a spatially correlated noise environment.

%outline
The rest of the paper is organized as follows. The system model is introduced in Section \ref{sec:Model} including a brief discussion of cyclostationarity. The proposed algorithm is detailed in Section \ref{sec:Proposed} including analysis of its detection performance under both AWGN and flat fading. Numerical results for various scenarios are presented in Section \ref{sec:Results}. Finally, the paper is concluded in Section \ref{sec:Conclusion}.

%notation
\textit{Notation:} $\left|\mathbf{A}\right|$ and $\rm{tr}(\mathbf{A})$ denote the determinant and trace of square matrix $\mathbf{A}$ respectively. $\mathbf{B}_{ij}$ denotes the $(i,j)$th element of the matrix $\mathbf{B}$ and $\mathbf{I}$ is the identity matrix. The superscripts $*$ and $H$ denote the complex conjugate and the Hermitian (conjugate transpose) operations, respectively. Given two random vectors $\mathbf{x}$ and $\mathbf{y}$, we define $\cov(\mathbf{x},\mathbf{y}) \triangleq E\{\mathbf{x}\mathbf{y}^H\}-E\{\mathbf{x}\}E\{\mathbf{y}^H\}$. Given column vector $\mathbf{x}$, $\diag\{\mathbf{x}\}$ denotes a square matrix with elements of $\mathbf{x}$ along its main diagonal and zeros everywhere else. We will use the notation $\mathcal{N}_c(\mathbf{m},\mathbf{\Sigma})$ to denote a proper (circularly symmetric) complex multivariate Gaussian distribution with mean $\mathbf{m}$ and covariance $\mathbf{\Sigma}$. Finally we use the notation $y=\mathcal{O}(g(x))$ to indicate that there exists some finite real number $b>0$ such that $\lim_{x\rightarrow\infty}\vert y/g(x)\vert\leq b$.

%%%%%%%%%%%%%%%%%%%%%%%%%%%%%%%%%%%%%%%%%%%%%%%%%%%%%%%%%%%%%%%%%%%%%%%%%%%%%%%
%%%%%%%%%%%%%%%%%%%%%%%%%%%%%%%%%%%%%%%%%%%%%%%%%%%%%%%%%%%%%%%%%%%%%%%%%%%%%%%

\section{Background and System Model}
\label{sec:Model}

\subsection{Background on Cyclostationarity}

A signal is considered to be cyclostationary if its statistical properties are periodic. Equivalently, if the cyclic autocorrelation function, defined as:
\begin{equation}
\label{eqn:CyclicCovariance}
R_{x}^{\alpha}(\tau)=\!\!\!\lim_{\Delta t\rightarrow\infty}\frac{1}{\Delta t}\!\int_{-\frac{\Delta t}{2}}^{\frac{\Delta t}{2}}\!\!x\!\left(t+\frac{\tau}{2}\right)\!x^{*}\!\left(t-\frac{\tau}{2}\right)\!e^{-j2\pi\alpha t}dt,
\end{equation}
is non-zero with some $\tau$ for at least one $\alpha \neq 0$, the signal is said to exhibit second-order cyclostationary property with $\alpha$ referred to as the cyclic frequency.

For example, in BPSK signals, cyclostationary features exist at $\alpha=\frac{k}{T_b}$ and at $\alpha=\pm 2 f_c + \frac{k}{T_b}$, where $T_b$ is the symbol period, $f_c$ is the carrier frequency, and $k\in\mathbb{Z}$. Detailed analysis of the cyclostationary features for various digital modulations can be found in \cite{Gardner1987a}.

\subsection{Signal Model and Assumptions}
We adopt a similar signal model as that used in \cite{Jitvanichphaibool2010a}. The spectrum sensing problem is to decide between two hypotheses: $\mathcal{H}_0$, where the signal is absent; and $\mathcal{H}_1$, where it is present. The received signal samples under the two hypothesis are given respectively as follows:
\begin{equation}
x(n)=\begin{cases}
\eta(n), & \mathcal{H}_0\\
s(n) + \eta(n), & \mathcal{H}_1.
\end{cases}
\end{equation}

The received signal, sampled at a rate of $1/T_s$, forms $M$ streams coming from each antenna with $N$ samples each. This received signal is defined as $
\mathbf{x}(n)\triangleq[x_{1}(n),x_{2}(n),\ldots,x_{M}(n)]^{T}$. The received signal is the superposition of $P$ signal sources (including both the SOI and any interferer) and can be expressed in vector form as
\begin{equation}
\label{eqn:ReceivedDecompose}
\mathbf{x}\left(n\right)=\sum_{j=1}^{P}\mathbf{h}_{j}\left(n\right)\otimes s_{j}\left(n\right)+\boldsymbol{\eta}\left(n\right),
\end{equation}
where $\otimes$ is the convolution operation over $n$ and $\boldsymbol{\eta}(n)$ is the receiver noise denoted by $\boldsymbol{\eta}(n)
\triangleq[\eta_{1}(n),\eta_{2}(n),\ldots,\eta_{M}(n)]^{T}$, where every $\eta_i$ is a purely stationary Gaussian random process ($R_{\eta}^{\alpha}(\tau)=0$ for any $\alpha\neq 0$) with variance of $\sigma^2_\eta$. For simplicity, we restrict that only one PU transmission, $s_1(n)$, is considered a SOI and that it is cyclostationary with a unique cyclic frequency $\alpha=\alpha_0$. The channel experienced by each of the $P$ sources is given by $\mathbf{h}_{j}(n)\triangleq[h_{j1}(n),h_{j2}(n),\ldots,h_{jM}(n)]^{T}$, where $h_{jk}(n)$ is the channel between the $j$th source and the $k$th antenna. We assume that the channel, although unknown to the receiver, stays constant over the spectrum sensing interval. Subsequently, we define the average signal-to-noise ratio to be
\begin{equation}
\label{eqn:snr}
{\rm SNR\triangleq}\frac{E\left\{ \mathbf{h}^{H}\mathbf{h}\right\} }{E\left\{ \boldsymbol{\eta}^{H}\boldsymbol{\eta}\right\}}.
\end{equation}

\subsection{Spatially Correlated Noise Environments}
\label{subsubsec:spatialcorr}
In the case of spatially correlated noise, which can happen when there is substantial ambient noise in the band, following \cite{Shrimpton1997}, we model $\boldsymbol{\eta}(n)$ to have a covariance matrix  given by $\mathbf{R}_{\boldsymbol{\eta\eta}}=\cov\{\boldsymbol{\eta},\boldsymbol{\eta}\}$ where
\begin{equation}
\{\mathbf{R}_{\boldsymbol{\eta\eta}}\}_{ij}=E\{\boldsymbol{\eta}^H_i\boldsymbol{\eta}_j\}=\begin{cases}
\sigma_\eta, & i=j\\
\sigma_\eta\rho_s^{\left\vert i-j \right\vert}, & i\neq j.
\end{cases}
\end{equation}
Thus with $\rho_s=0$, the covariance matrix simplifies to $\sigma_\eta\mathbf{I}$ giving spatially white noise, while $\rho_s=1$ gives fully correlated noise over all antennas. Varying degrees of partial correlation can be achieved by setting $0<\rho_s<1$.

%%%%%%%%%%%%%%%%%%%%%%%%%%%%%%%%%%%%%%%%%%%%%%%%%%%%%%%%%%%%%%%%%%%%%%%%%%%%%%%
%%%%%%%%%%%%%%%%%%%%%%%%%%%%%%%%%%%%%%%%%%%%%%%%%%%%%%%%%%%%%%%%%%%%%%%%%%%%%%%

\section{Proposed Algorithm}
\label{sec:Proposed}
In this section, we describe the proposed method. We focus on a single cycle frequency detection, but this approach could be generalized to multi-cycle detection.

\subsection{Canonical Analysis}
The key idea of this detection algorithm is based on the theory of canonical analysis. As discussed in \cite{Lawley1959}, and subsequently utilized in \cite{Schell1990a}, the number of common factors between two $M\times1$ time-series vectors $\mathbf{x}(n)$ and $\mathbf{y}(n)$ can be inferred from the rank of the matrix
\begin{equation}
\label{eqn:CommonFactors}
\mathbf{R}=\mathbf{R}_{\mathbf{xx}}^{-1}\mathbf{R}_{\mathbf{xy}}^{}\mathbf{R}_{\mathbf{yy}}^{-1}\mathbf{R}_{\mathbf{yx}}^{},
\end{equation}
where we define $\mathbf{R}_{\mathbf{x}\mathbf{y}}\triangleq\cov(\mathbf{x},\mathbf{y})$ given the two random vectors $\mathbf{x}$ and $\mathbf{y}$.

Canonical analysis (see \cite{Anderson2003} for details) aims to find the relationships between two groups of variables in a data set. Given two random vectors $\mathbf{x}$ and $\mathbf{y}$, of length $m$ and $n$ respectively, canonical analysis aims to find at most $\min\{m,n\}$ pairs of $(\mathbf{u}_i,\mathbf{v}_i)$ such that the correlation between the linear combinations, $U_i\triangleq\mathbf{u}_i^H\mathbf{x}$ and $V_i\triangleq\mathbf{v}_i^H\mathbf{y}$, is maximized. An additional restriction is that $U_i$ and $V_i$ must be uncorrelated with $U_j$ and $V_j$ for $i \neq j$. These linear combinations are referred to as canonical variates. The canonical variates are sorted in decreasing order of correlation such that the first canonical variates, $U_1$ and $V_1$, have the highest correlation. The correlation coefficient, $\rho_i$, between $U_i$ and $V_i$ is referred to as the $i$th canonical correlation. This procedure of finding $\mathbf{u}_i$ and $\mathbf{v}_i$ can be efficiently performed using a singular value decomposition (SVD) and the square of the canonical correlations can be found by finding the eigenvalues of (\ref{eqn:CommonFactors}).

In the context of cyclostationary spectrum spectrum sensing, canonical analysis provides us with a very powerful tool to optimally combine the signals from $M$ antennas and find the canonical correlations, $\rho_i$, resulting from up to $M$ mutually uncorrelated cyclostationary signals. This can be accomplished using the Cyclic Correlation Significance Test (CCST) \cite{Schell1990a} by performing canonical analysis on $\mathbf{x}(n)$ and $\mathbf{x}(n-\tau)e^{-j2\pi\alpha nT_s}$ for a given lag $\tau$ and cyclic frequency $\alpha$. By finding the canonical correlations between these two sets of data, we are in effect measuring the maximum amount of cyclic correlation for all possible linear combinations of the signals coming from the $M$ antennas. A threshold can then be applied on the combined $\rho_i$'s to determine the presence or absence of the PU. Additionally, some cyclic frequencies, such as those located on $\alpha=\pm 2f_c$ for BPSK, only appear in the conjugate cyclic correlation. These can also be detected by instead performing the canonical analysis with $\mathbf{x}^*(n-\tau)e^{-j2\pi\alpha nT_s}$.

Prior to performing the detection, we pick the lag $\tau$ that provides the best detection performance based on the modulation format used by the PU. This could be done off-line by performing the maximization, $\tau_0=\arg\max_\tau |R_s^{\alpha_0}(\tau)|$.

\subsection{Algorithm Description}

The steps of the algorithm are summarized as follows:
\begin{enumerate}
	\item Estimate the covariance matrix of size $M\times M$
	\begin{equation}
	\label{eqn:Covariance}
	\hat{\mathbf{R}}_{\mathbf{xx}}(\tau_0)=\frac{1}{N-1-\tau_0}\sum_{n=0}^{N-1-\tau_0}\mathbf{x}\left(n\right)\mathbf{x}^{H}\left(n-\tau_0\right).
	\end{equation}
	\item Estimate the cyclic correlation matrix using a cyclic cross-correlogram at cyclic frequency $\alpha_0$ and lag $\tau_0$, defined as
	\begin{equation}
	\label{eqn:CyclicCovariance}
	\hat{\mathbf{R}}_{\mathbf{xx}}^{\alpha_0}(\tau_0)=\frac{1}{N-1-\tau_0}\!\!\sum_{n=0}^{N-1-\tau_0}\!\!\!\!\!\mathbf{x}\left(n\right)\mathbf{x}^{H}\!\!\left(n-\tau_0\right)e^{-j2\pi\alpha_0 nT_s}.
	\end{equation}
	We will refer to the $\tau_0$-lag covariance matrices for both conventional and cyclic autocorrelation function simply as $\hat{\mathbf{R}}_{\mathbf{xx}}$ and $\hat{\mathbf{R}}_{\mathbf{xx}}^{\alpha_0}$ from this point for the sake of brevity, since other $\tau$ are not utilized by the proposed algorithm. The dependence on $\tau$ will be indicated explicitly whenever necessary. The CCST is then calculated by finding the matrix
	\begin{equation}
	\label{eqn:Metric}
	\hat{\mathbf{R}}=\hat{\mathbf{R}}_{\mathbf{xx}}^{-1}\hat{\mathbf{R}}_{\mathbf{xx}}^{\alpha_0} \hat{\mathbf{R}}_{\mathbf{xx}}^{-1}\hat{\mathbf{R}}_{\mathbf{xx}}^{\alpha_0 H}.
	\end{equation}
	\item Find the eigenvalues, $\boldsymbol{\mu}=\left[\mu_1^2,\mu_2^2,\ldots,\mu_M^2\right]^{T}$, of $\mathbf{\hat{R}}$.
	\item Compute the test statistic by combining the eigenvalues using
	\begin{equation}
	\label{eqn:lambda}
	\lambda \triangleq \prod_{i=1}^{M}\left(1-\mu_{i}^2\right),
	\end{equation}
	and finally calculate the test statistic:
	\begin{equation}
	\label{eqn:TestStatistic}
	\mathcal{T^{\alpha}_{\mathbf{xx}}}\triangleq -m\ln\lambda.
	\end{equation}
	The factor $m\triangleq N-M-1$ is used to scale the test statistic so that its distribution is independent of the number of samples used \cite[Sec. 8]{Bartlett1938}.
	\item Decision: 
	$\mathcal{T^{\alpha}_{\mathbf{xx}}}\mathop{\gtrless}_{\mathcal{H}_0}^{\mathcal{H}_1}\gamma$,
	where $\gamma>0$ is a threshold chosen to achieve constant false alarm rate (CFAR) which will be discussed in the following section.
	
\end{enumerate}

Note that all $\mathbf{x}^H(n)$ can be replaced with $\mathbf{x}^T(n)$ if the conjugate cyclic correlation matrix is needed. We refer to the version of the algorithm that uses $\mathbf{x}^H(n)$ as the non-conjugate cyclic correlation significance test (NC-CCST) while the other is the conjugate cyclic correlation significance test (C-CCST). For the test statistic of each, we will use the notations $\mathcal{T^{\alpha}_{\mathbf{xx}}}$ and $\mathcal{T^{\alpha}_{\mathbf{xx^*}}}$ respectively.

\subsection{Distribution Under $\mathcal{H}_0$ and Constant False-Alarm Rate}
\label{subsec:h0}
Two key parameters are used to evaluate the performance of spectrum sensing algorithms. The detection probability or $P_{\rm{D}}$ is the probability of being at $\mathcal{H}_1$ and accurately detecting the PU ($P_{\rm{D}}\triangleq\Pr(\mathcal{T^{\alpha}_{\mathbf{xx}}}>\gamma \mid \mathcal{H}_1)$). On the other hand, the false alarm probability, $P_{\rm{FA}}$, is the probability of being at $\mathcal{H}_0$ and mistakenly detecting a PU ($P_{\rm{FA}}\triangleq\Pr(\mathcal{T^{\alpha}_{\mathbf{xx}}}>\gamma \mid \mathcal{H}_0)$).

It has been shown in \cite[Sec. 8]{Bartlett1938} that the limiting distribution ($N\rightarrow\infty$) of the test statistic (\ref{eqn:TestStatistic}) for real and normally distributed random vectors approaches a $\chi^2$ distribution with degree-of-freedom $M^2$. Following a similar proof, it can also be shown that for zero mean, complex Gaussian random variables, the distribution is also $\chi^2$ with degree-of-freedom $M^2$ when using the NC-CCST and $M(M+1)$ for the C-CCST.

Based on the distribution of $\mathcal{T^{\alpha}_{\mathbf{xx}}}$ under $\mathcal{H}_0$, the detection threshold $\gamma$ can be set to achieve a desired $P_{\rm{FA}}$ by satisfying
\begin{equation}
\label{eqn:threshold}
\int_{\gamma}^{\infty}f_{\chi_{k}^{2}}\left(x\right)dx=P_{\rm{FA}},
\end{equation}
where
\begin{equation}
\\k=\begin{cases}
M^{2} & \text{NC-CCST}\\
M(M+1) & \text{C-CCST}
\end{cases},
\end{equation}
and $f_{\chi_{k}^{2}}(\cdot)$ is the probability density function (pdf) of a $\chi^2$ random variable with degree-of-freedom $k$.

\begin{figure}
\centering\includegraphics[width=21pc]{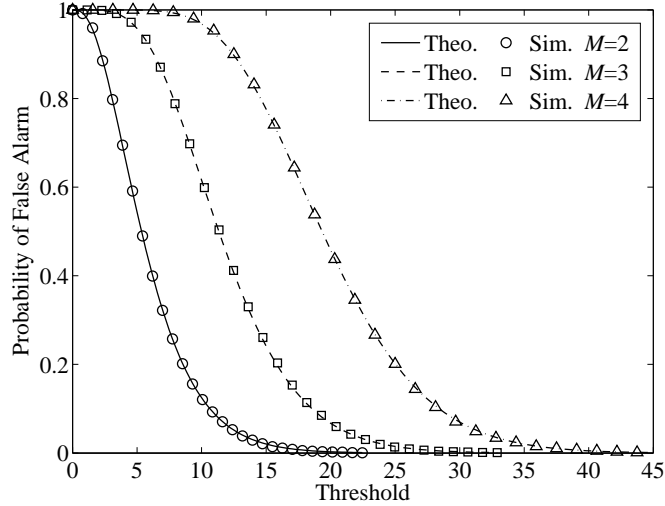}
\caption{Verification of the asymptotic distribution of the proposed test statistic (C-CCST) under $\mathcal{H}_0$ with different number of antennas ($M=\{2,3,4\}$). These plots show the accuracy of the analytical expression under $N=1000$ number of samples per antenna (SNR=-10 dB).}
\label{fig:pdf}
\end{figure}

These asymptotic distributions are verified to closely match simulation in Fig.~\ref{fig:pdf} for $N=4000$. Due to the scaling factor in (\ref{eqn:TestStatistic}), the distribution is independent of $N$. The empirical pdfs for two different $\sigma^2_\eta$ values are also shown to demonstrate how the test statistic's distribution under $\mathcal{H}_0$ is independent of noise power.

As introduced in Section~\ref{subsubsec:spatialcorr}, spatially correlated noise happens whenever $\mathbf{R}_{\boldsymbol{\eta\eta}}$ is non-diagonal. This could also be interpreted as having a transformed noise vector
\begin{equation}
\label{eqn:noiselin}
\boldsymbol{\eta}'=\mathbf{A}\boldsymbol{\eta}\quad\rm{s.t.}\quad \sigma_{\eta}\mathbf{AA}^H=\mathbf{R}_{\boldsymbol{\eta}'\boldsymbol{\eta}'},
\end{equation}
where $\mathbf{A}$ is an $M\times M$ matrix that determines the spatial correlation among antennas. In order to see the effect of correlated noise on the distribution of $\mathcal{T}^{\alpha}_{\mathbf{xx}}$ under $\mathcal{H}_0$, it is helpful to use an alternate interpretation of canonical correlation as given in \cite[Eqn. 2.1]{Sugiura1969} such that
\begin{equation}
\label{eqn:cclr}
\lambda = \frac{\begin{vmatrix}
      \mathbf{\hat{R}^{}_{xx}} & \mathbf{\hat{R}_{xx}}^{\alpha_0} \\
      \mathbf{\hat{R}_{xx}}^{\alpha_{0}H} & \mathbf{\hat{R}_{xx}} ^H
    \end{vmatrix}}{\left\vert\mathbf{\hat{R}_{xx}}\right\vert^2},
\end{equation}
using the covariance matrix estimates given in (\ref{eqn:Covariance}) and (\ref{eqn:CyclicCovariance}) and $\lambda$ as defined in (\ref{eqn:lambda}). Using (\ref{eqn:noiselin}), in conjunction with the multiplication property of determinants, and determinants of block diagonal matrices, we find the value of $\lambda$ for correlated noise ($\lambda_c$) to be
\begin{eqnarray}
\lambda_{c} &=& \frac{\left\vert
\begin{bmatrix}
      \mathbf{A} & \mathbf{0} \\
      \mathbf{0} & \mathbf{A}
\end{bmatrix}
\begin{bmatrix}
      \mathbf{\hat{R}_{xx}} & \mathbf{\hat{R}_{xx}}^{\alpha_0} \\
      \mathbf{\hat{R}_{xx}}^{\alpha_{0}H} & \mathbf{\hat{R}_{xx}} ^H
\end{bmatrix}
\begin{bmatrix}
      \mathbf{A} & \mathbf{0} \\
      \mathbf{0} & \mathbf{A}
\end{bmatrix}^H
\right\vert}{\left\vert\mathbf{A\hat{R}^{}_{xx}}\mathbf{A}^H\right\vert^2}\\
&=&\frac{\left\vert\mathbf{A}\right\vert^4
\begin{vmatrix}
      \mathbf{\hat{R}_{xx}} & \mathbf{\hat{R}_{xx}}^{\alpha_0} \\
      \mathbf{\hat{R}_{xx}}^{\alpha_{0}H} & \mathbf{\hat{R}_{xx}} ^H
\end{vmatrix}}
{\left\vert\mathbf{A}\right\vert ^4\left\vert\mathbf{\hat{R}_{xx}}\right\vert^2}=\lambda.
\end{eqnarray}
Therefore the test statistic under $\mathcal{H}_0$ is invariant to any linear transformation on the noise measurements. As such the same expression for $P_{\rm{FA}}$ as well as the threshold for maintaining constant $P_{\rm{FA}}$ is applicable even for correlated noise environments.

\subsection{Distribution Under $\mathcal{H}_1$ and Probability of Detection}

In this section we derive the distribution of the proposed test statistic, $\mathcal{T^{\alpha}_{\mathbf{xx}}}$, under $\mathcal{H}_1$. We begin by summarizing the prior work in statistics leading to the derivation of the complete non-null distribution of the canonical correlations. We show how this result is parameterized by the canonical correlation, $\rho$, between the two signals being tested. In the context of cyclostationary spectrum sensing, this corresponds to the value of the signal's cyclic autocorrelation, $R_{ss}^\alpha \triangleq  E\left[s(n)s^*(n)e^{-j2\pi\alpha_0 nT_s}\right]$. We then derive the canonical correlation, $\rho$, resulting from $ R_{ss}^\alpha$ with a given $\sigma_\eta$, $N$, $
\mathbf{R}_{\boldsymbol{\eta}\boldsymbol{\eta}}$ and $\mathbf{h}$. Using this canonical correlation, the complete distribution of the test statistic under $\mathcal{H}_1$ is derived. Once the distribution is found, it allows us to find the theoretical probability of detection given as:
\begin{equation}
P_{\rm{D}} =\Pr(\mathcal{T^{\alpha}_{\mathbf{xx}}}>\gamma \mid \mathcal{H}_1)=1-F_{\mathcal{T^{\alpha}_{\mathbf{xx}}}|\mathcal{H}_1}\left(\gamma\right),
\end{equation}
where $F_{\mathcal{T^{\alpha}_{\mathbf{xx}}}|\mathcal{H}_1}\left(\cdot\right)$ is the cdf of the test statistic under $\mathcal{H}_1$ which we find in the rest of this subsection.

Several works in the past have contributed to deriving the non-null distribution of the test of significance of the canonical correlations originally proposed by Bartlett \cite{Bartlett1947}. In \cite{Lawley1959}, the approximate means and variances of the highest eigenvalue were found in the asymptotic case with only one non-zero eigenvalue ($\mu_1\neq0, \mu_i=0, i>1$ ) and normality assumption. The distribution of the actual likelihood-ratio criteria, of the same form as (\ref{eqn:TestStatistic}), was found in \cite{Sugiura1969}. However, both of these results break down in the case of local alternatives, which correspond to the cases when the distribution under $\mathcal{H}_1$ is very close to the null hypothesis. In relation to the CCST, this corresponds to having a very-low SNR, which is clearly the case of interest in the spectrum sensing problem.

Finally, the non-null distribution of the likelihood ratio criteria for covariance matrix under local alternatives were found in \cite{Sugiura1973}. This criteria is used to test the independence between two multivariate random variables by combining the canonical correlation into a single test statistic. We use the same criteria to test for independence between the signal of interest and a frequency shifted version of itself. After the publication of this distribution, several works have focused on eliminating the normality assumption \cite{Muirhead1980,Ogasawara2007}. However, in the case of spectrum sensing for CR, it is more likely to deal with detection under very-low SNR. If the noise is assumed AWGN and the $\sigma_\eta\geq\sigma_s$, the received signal, $\mathbf{x}(n)$, is approximately normal and the results for canonical analysis assuming normality can be used.

Following \cite[Thm. 4.1]{Sugiura1973}, the cdf of the the test statistic under local alternatives is found to be asymptotically distributed as a non-central chi-square. We have,
\begin{equation}
\label{eqn:distrib}
F_{\mathcal{T^{\alpha}_{\mathbf{xx}^*}}|\mathcal{H}_1}\left(x\right)=F_{\chi'^2}(x,M(M+1),\delta^2) + \mathcal{O}(m^{-1}),
\end{equation}
where $F_{\chi'^2}(x,d,\delta^2)$ is the cdf of the non-central chi-square random variable with degree-of-freedom of $d$ and non-centrality parameter $\delta^2$. The actual value of the non-centrality parameter can be found as $\delta^2=\rm{tr}(\Theta^2)$, where $\Theta=\sqrt{m}\diag\{\rho_1,\rho_2,\ldots,\rho_M\}$. Where, $\rho^2_i$ for $1<i<M$ are the eigenvalues of (\ref{eqn:Metric}) using ensemble averages. Thus, they are solutions to the equation
\begin{equation}
\left\vert \mathbf{\hat{R}}_{\mathbf{xx}}^{-1}\mathbf{\hat{R}}_{\mathbf{xx}}^{\alpha_{0}}\mathbf{\hat{R}}_{\mathbf{xx}}^{-1}\mathbf{\hat{R}}_{\mathbf{xx}}^{\alpha_{0}H}-\rho_{i}^{2}\mathbf{I}\right\vert =0. 
\end{equation}

The true distribution deviates from the non-central chi-square with lower number of samples as indicated by the additional $\mathcal{O}(m^{-1})$ term in (\ref{eqn:distrib}). The complete distribution up to the order $\mathcal{O}(m^{-2})$ can be found in \cite[Thm. 4.1]{Sugiura1973} which for the sake of brevity is no longer presented here. This more accurate distribution is an expansion based on non-central chi-square random variables of higher degrees-of-freedom with non-centrality parameters of $\rm{tr}(\Theta^4)$ and $\rm{tr}(\Theta^6)$. As such, a very accurate expression for $F_{\mathcal{T^{\alpha}_{\mathbf{xx}}}|\mathcal{H}_1}\left(x\right)$ can be calculated with knowledge of $\Theta$. 

Let's assume, for simplicity, that only one signal of interest $s(n)$ has cyclic frequency $\alpha$. Therefore, $\rho_i=0$ for $i>1$. This assumption applies in almost all cases since two communication signals will, with high probability, have different cyclic frequencies. However, it is straightforward to extend these results to the case of multiple signals with exactly the same $\alpha$, since this simply corresponds to additional non-zero $\rho_i$ and the derivations presented here are still applicable. Thus, these scenarios can be treated theoretically as being single signal scenarios as long as the number of signals is less than or equal to the number of antennas $M$.

The channel, $\mathbf{h}$, is also assumed to be constant over one sensing period. Therefore, the expression for the distribution derived in this subsection is for a particular channel instance. Later we extend these results for flat-fading channels by integrating the distributions over the statistics of the fading channel.

With these assumptions, we proceed with finding the values of $\rho_i$ as a function of the channel $\mathbf{h}$, the signal of interest $s(n)$ and the noise covariance $\mathbf{R}_{\boldsymbol{\eta\eta}}$. The received signal can be expressed as $\mathbf{x}(n) = s(n)\mathbf{h} + \boldsymbol{\eta}(n)$. Using our initial assumption that the signal has unit power and is uncorrelated with the noise, the asymptotic zero-lag covariance matrix then becomes
\begin{eqnarray}
\mathbf{R}_{\mathbf{xx}}&=&\cov\left(\mathbf{x}(n),\mathbf{x}(n)\right)\\
&=&\mathbf{h}\mathbf{h}^H E\left[s(n)s^*(n)\right] +\mathbf{R}_{\boldsymbol{\eta\eta}}\\
\label{eqn:rxx}&=&\mathbf{h}\mathbf{h}^H+\mathbf{R}_{\boldsymbol{\eta\eta}}.%\sigma_{\eta}^2\mathbf{I}_M,
\end{eqnarray}
On the other hand, recalling that noise has no cyclic features, the asymptotic cyclic cross covariance matrix can be found as
\begin{eqnarray}
\mathbf{R}_\mathbf{xx}^{\alpha_0}&=&\cov\left\{\mathbf{x}\left(n\right),\mathbf{x}(n)e^{-j2\pi\alpha_0 nT_s}\right\}\\
&=&\mathbf{h}\mathbf{h}^H E\left[s(n)s^*(n)e^{-j2\pi\alpha_0 nT_s}\right]\\
\label{eqn:rxxalpha}&=&\mathbf{h}\mathbf{h}^H R_{ss}^\alpha.
\end{eqnarray}
Subsequently the conjugate version, $R_{ss^*}^\alpha$, could also be used by replacing $s^*(n)$ with $s(n)$.

We again use the alternate interpretation in (\ref{eqn:cclr}) to find that the canonical correlation under $\mathcal{H}_1$ as
\begin{equation}
\label{corrcoeff}
\rho = \frac{\left\Vert \mathbf{A}^{-1}\mathbf{h} \right\Vert^2 \left\vert R_{ss}^\alpha\right\vert}{\left\Vert \mathbf{A}^{-1}\mathbf{h} \right\Vert^2+\sigma^2_\eta},
\end{equation}
where the noise covariance matrix is $\mathbf{R}_{\boldsymbol{\eta}\boldsymbol{\eta}}=\sigma_\eta \mathbf{A}\mathbf{A}^H$. The details of the derivation of (\ref{corrcoeff}) are provided in Appendix \ref{app:Rho}. In the case of spatially uncorrelated noise, we have $\mathbf{A}=\mathbf{I}_M$ so that the true correlation becomes
\begin{equation}
\label{corrcoeffrho0}
\rho = \frac{\left\Vert \mathbf{h} \right\Vert^2 \left\vert R_{ss}^\alpha\right\vert}{\left\Vert \mathbf{h} \right\Vert^2+\sigma^2_\eta},
\end{equation}
which is only dependent on the 2-norm of the channel coefficient, the noise variance $\sigma_\eta$, and the cyclic correlation for the chosen $\alpha$. A derivation of $R_{ss}^\alpha$ is given in \cite{Rebeiz2011} for various modulation schemes including BPSK, MSK, and QAM.

\begin{figure}
\centering\includegraphics[width=21pc]{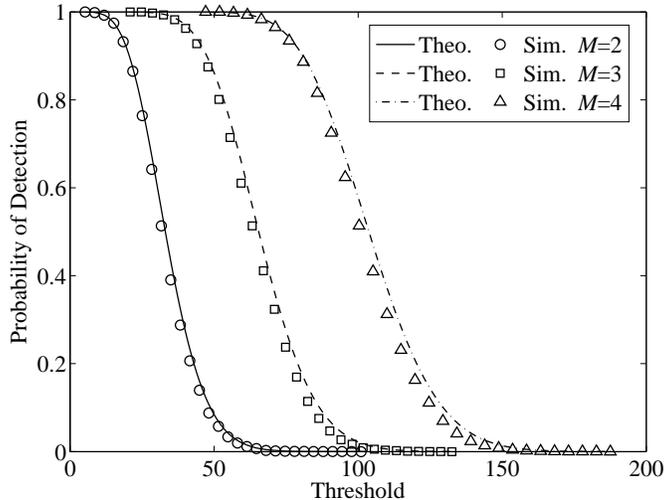}
\caption{Verification of the asymptotic distribution of the proposed test statistic (C-CCST) under $\mathcal{H}_1$ with different number of antennas ($M=\{2,3,4\}$). These plots show the accuracy of the analytical expression under $N=1000$ number of samples per antenna (SNR=-10 dB).}
\label{fig:h1pdf}
\end{figure}

A similar derivation can be done for C-CCST by replacing $R_{ss}^\alpha$ with $R_{ss^*}^\alpha$. Thus for a single SOI scenario we have
\begin{equation}
\Theta=\sqrt{m}\left[\rho,0,\ldots,0\right]\mathbf{I}_M.
\end{equation}
Which when substituted to (\ref{eqn:distrib}) gives us the complete distribution. In Fig.~\ref{fig:h1pdf}, we show that the theoretical results are in very close agreement to the simulation.

\subsection{Rayleigh Fading}
\label{subsec:rayleighfading}
Now that we have the complete distribution for both the null and non-null hypothesis with AWGN parameterized by a particular instance of the channel $\mathbf{h}$ and the noise covariance $\mathbf{R}_{\boldsymbol{\eta}\boldsymbol{\eta}}$, we can proceed to evaluate the performance of the proposed algorithm under a flat-fading environment.

In particular we use a flat fading channel model where the channel vector, $\mathbf{h}$, remains constant during the whole frame of samples used in detection. This is described using a channel vector for the $i$th frame as $\mathbf{h}_i=[r_1e^{j\theta_1},r_2e^{j\theta_2},\ldots,r_Me^{j\theta_M}]^{T}$,
where $r_n$ is a Rayleigh distributed random variable of unit variance and $\theta_n$ is a uniformly distributed random variable in $[0,2\pi]$.

We first assume spatially uncorrelated noise ($\mathbf{R}_{\boldsymbol{\eta}\boldsymbol{\eta}}=\sigma_\eta\mathbf{I}$). Based on (\ref{corrcoeffrho0}), the true correlation coefficient, which eventually determines the non-centrality parameter in (\ref{eqn:distrib}) is a function of $\left\Vert \mathbf{h} \right\Vert^2=\mathbf{h}^H\mathbf{h}$. The distribution of $\mathbf{h}$ can be seen as a complex normal random vector of size $M\times1$ which results in $\left\Vert \mathbf{h} \right\Vert^2\sim\chi^2_{2M}$. Combining (\ref{eqn:distrib}) and (\ref{corrcoeffrho0}), we find the new cdf of our test statistic under Rayleigh flat-fading and spatially uncorrelated noise as:
\begin{equation}
\label{eqn:rayleigh}
	F^*_{\mathcal{T^{\alpha}_{\mathbf{xx}}}|\mathcal{H}_1}\left(x\right)=\int_0^\infty F_{\chi'^2_{M(M+1)}}\left(x,\frac{\beta\left|R_{ss}^\alpha(\tau)\right|}{\beta+\sigma_\eta^2}\right)
	f_{\chi^2_{2M}}(\beta)d\beta.
\end{equation}

In the case of spatially correlated noise, the random variable $\beta=\left\Vert\mathbf{A}^{-1}\mathbf{h}\right\Vert^2=\mathbf{h}^H\left(\mathbf{A}\mathbf{A}^H\right)^{-1}\mathbf{h}$ becomes a generalized $\chi^2$ r.v. instead.

Although a closed form expression is very difficult to derive for such an expression due to the presence of the non-central chi-square, there is still some insight to be gained by numerically integrating (\ref{eqn:rayleigh}) to arrive at $P_{\rm{D}}$ and $P_{\rm{FA}}$ expressions under flat fading.

\subsection{Comparison With Existing Approaches}
\label{subsec:others}
The algorithms for multiple-antenna spectrum sensing based on cyclostationarity that are currently in the literature can generally be classified into two categories. The simplest method to do this is to find the sum of the spectral correlation test statistic estimated individually from each antenna \cite{Sadeghi2008}. We refer to this approach as SUM-MSDF (where MSDF means Modified Spectral Density Function). The MSDF is defined as the spectral correlation function (SCF) normalized by signal energy as discussed in \cite{Jitvanichphaibool2010a}.

Another existing approach is to sum the raw samples from each antenna and then perform a single spectral correlation test. However, we encounter a problem when the channel is not simply AWGN but instead has random fading. In this case, each antenna will have some unknown phase offset and attenuation. Thus, simply adding the raw samples non-coherently would decrease the probability of detection. This problem is addressed in \cite{Chen2008} by first eliminating the phase rotation of signal samples coming from each antenna. An estimate of the relative phase difference between each antenna is calculated by finding both the cyclic correlation of one antenna chosen as reference (auto-spectral correlation function or auto-SCF) and the cross-cyclic correlation of every other antenna and the reference antenna. The phase difference can then be extracted from these two. We refer to this method in our comparisons as Equal Gain Combining (EGC).

Finally, Maximal Ratio Combining (MRC) is used in \cite{Jitvanichphaibool2010a}. Blind channel estimation is achieved by taking the vector corresponding to the highest singular value of (\ref{eqn:CyclicCovariance}) as an estimate of the channel, $\hat{\mathbf{h}}$. The raw samples from each antenna are combined using
\begin{equation}
	y(n) = \frac{\hat{\mathbf{h}}^H\mathbf{x}(n)}{\Vert \hat{\mathbf{h}}\Vert } .
\end{equation}
The cyclic correlation test is then performed on the combined samples $y(n)$. This method is called MSDF with blind maximal ratio combining or BMRC-MSDF. It was shown to outperform the other techniques but at the cost of additional complexity due to the channel estimation and combining. One issue with this approach is the fact that the cyclic correlation is calculated twice. The first is used to blindly estimate the channel and the second to perform the detection on the combined samples. In contrast, the method proposed in this paper only needs to perform the first part of BMRC-MSDF, finding the eigenvalues, and then uses the eigenvalues themselves to infer the presence or absence of the PU.

\subsection{Advantages of the Proposed Algorithm}

As with other cyclostationarity-based spectrum sensing algorithms, one major advantage of the proposed method is its robustness to the noise uncertainty problem. Since the noise is assumed to be stationary and does not exhibit cyclostationarity at any $\alpha \neq 0$, its cyclic correlation approaches zero as $N\rightarrow\infty$. Thus, the effect of any error in the noise power estimate on the detection probability can be eliminated by taking more samples. However, in the interest of conserving power and arriving at a timely decision, both of which are high priority in the case of CR applications, we aim to minimize $N$ needed to achieve a target $P_{\rm{D}}$. This presents another, more subtle, issue related to noise uncertainty. 

In the non-asymptotic scenario, the methods based on the SCF (BMRC-MSDF, EGC and SUM-MSDF) under $\mathcal{H}_0$ have been shown to depend on both $N$ and the noise power $\sigma_\eta$ \cite{Rebeiz2011}. Therefore, the proper detection threshold is still a function of the noise variance. By incorrectly specifying this threshold, the detector could be at the wrong point in the receiver operating characteristic (ROC) curve. Equivalently, the target CFAR cannot be achieved. However, as previously discussed and demonstrated in Fig.~\ref{fig:pdf}, the proposed test statistic is independent of both $\sigma^2_\eta$ and $N$. Consequently, the threshold $\gamma$ only needs to be chosen once for a given number of antennas $M$ to guarantee CFAR. This property has been shown for other eigenvalue-based approaches \cite{Zeng2009}. It derives from the fact that noise power estimation is built-in to the algorithm.

\subsection{A Note on Complexity}
\label{subsec:complexity}
We provide an approximate complexity comparison of the proposed algorithm with the best performing existing algorithm (BMRC-MSDF) by taking number of complex multiplications required for each under the same number of samples $N$. Since the cyclic covariance operation and the SVD are common to both algorithms, they are not included in the analysis.

Assuming the MSDF is calculated using an $N_S$-point Fast Fourier Transform (FFT) it requires in the order of $N\log_2(N_S)$ multiplications. In addition, $(M+1)N$ multiplications are needed to perform the MRC and normalization. Finally, the correlation in frequency uses $NN_S/2$ multiplications. Thus, the BMRC-MSDF approach performs in the order of $N(\log_2(N_S)+N_S/2+M+1)$ multiplications without taking into account the SVD and the cyclic covariance.

In comparison, the proposed EV-CSS method finds the conventional covariance in addition to an eigenvalue decomposition (EVD) and the same cyclic covariance as BMRC-MSDF, or in the order of $NM^2$ multiplications. The operation $\hat{\mathbf{R}}_{\mathbf{xx}}^{-1}\hat{\mathbf{R}}_{\mathbf{xx}}^{\alpha}$ in (\ref{eqn:Metric}) is essentially the solution to a generalized linear system which can be seen as an LU decomposition requiring approximately $2M^3/3$ multiplications. Therefore, the EV-CSS approach requires in the order of $NM^2+2M^3/3$ multiplications in addition to the common operations with BMRC-MSDF. Since $M$ is typically much less than both $N$ and $N_S$, there is overall a significant decrease in complexity with the proposed algorithm. For example, if we take $N=4000$, $N_S=128$, and $M=2$, (same parameters used in \cite{Jitvanichphaibool2010a}), the BMRC-MSDF requires $\sim$296K multiplications while EV-CSS needs only $\sim$16K multiplications, without counting the common operations.

%%%%%%%%%%%%%%%%%%%%%%%%%%%%%%%%%%%%%%%%%%%%%%%%%%%%%%%%%%%%%%%%%%%%%%%%%%%%%%%
%%%%%%%%%%%%%%%%%%%%%%%%%%%%%%%%%%%%%%%%%%%%%%%%%%%%%%%%%%%%%%%%%%%%%%%%%%%%%%%

\section{Numerical Results and Discussion}
\label{sec:Results}

In this section, simulation results are presented in order to compare the performance of the proposed algorithm with the various existing techniques discussed in Section~\ref{subsec:others}. In addition, theoretical plots are included to further verify the analytical expressions of the proposed method's performance.

For these simulations we assume that only one PU has a feature at the chosen cyclic frequency, $\alpha_0$. This PU is assumed to be transmitting a BPSK signal at a carrier frequency $f_c=80$ KHz with symbol period of 25 $\rm\mu$s. Each antenna of the SU is sampled at a rate $f_s=320$ kHz. For all algorithms, the same cyclic frequency located at $\alpha_0=2f_c$ is used. This cyclic feature is only present in the conjugate cyclic autocorrelation which means the C-CCST statistic is used. This feature is chosen because it is the highest magnitude feature among all cyclic frequencies. The maximum cyclic autocorrelation at this cyclic frequency is observed at $\tau_0=0$ which is the lag we will be using for all EV-CSS simulations.

\subsection{Threshold Selection for EV-CSS}
One key advantage of the proposed EV-CSS scheme over other spectrum sensing schemes is the simplicity of threshold selection to achieve CFAR. As discussed in Section~\ref{subsec:h0} the detection threshold, $\gamma$, is only dependent on $M$ and not on $N$ or $\sigma_\eta$. As a result of this, in a practical implementation of EV-CSS, the threshold can be pre-calculated using only the $\chi^2$ distribution parameterized by the number of antennas. As we will show in the following subsections, the theoretically determined threshold achieves the desired $P_{\rm{FA}}$ and matches very well with simulations. Thus, only a single threshold needs to be stored. For the following simulations we set the CFAR to $P_{\rm{FA}}=0.1$ unless otherwise stated. 

\subsection{ROC and Detection Probability Versus SNR}

We first consider $M=2$ antennas in the SU. The channel between the PU and each antenna of the SU, is modeled as a quasi-static Rayleigh fading channel with channel  $\mathbf{h}$ as described in \ref{subsec:rayleighfading}. The fading is assumed to be frequency-flat and remains constant during the whole frame of $N=4000$ samples per antenna used for detection. The noise in the antennas is assumed to be distributed as a zero-mean circularly symmetric complex-Gaussian, $\boldsymbol{\eta}\sim\mathcal{N}_c(\mathbf{0},\sigma_\eta\mathbf{I})$. The PU signal energy is assumed to be unit energy and $\sigma_\eta$ is chosen to achieve an average SNR across antennas as defined in (\ref{eqn:snr}).

\begin{figure}
\centering
\includegraphics[width=21pc]{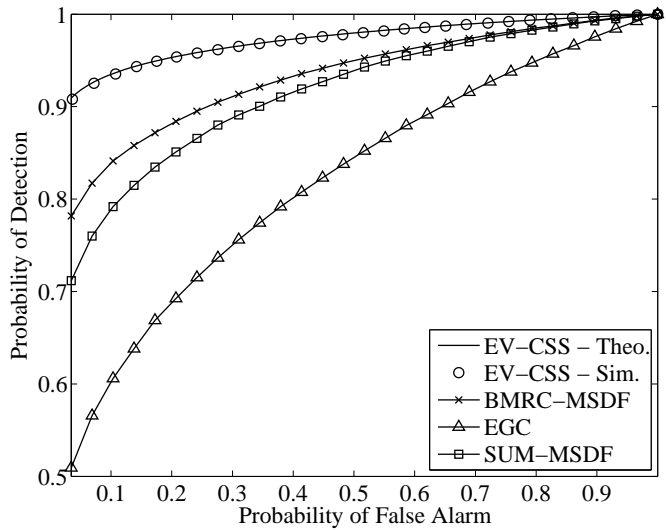}
\caption{Receiver Operating Characteristic (ROC) of different cyclostationary-based spectrum sensing algorithms under Rayleigh flat-fading (SNR=-10 dB, $N=4000$).}
\label{fig:roc_fading}
\end{figure}

Using these assumptions the ROC curves under SNR $=-10$ dB and $N=4000$ samples for the proposed algorithm and the other cyclic-based approaches are shown in Fig.~\ref{fig:roc_fading} for comparison. Both theoretical results and Monte Carlo simulations (with 50,000 trials) are shown for EV-CSS, while only the Monte Carlo simulations are shown for the other methods since these are already analyzed in the respective works that proposed them. We have verified that these results agree with simulations presented in these prior work under similar assumptions.

As seen in Fig.~\ref{fig:roc_fading}, there is very strong agreement in the theoretical and simulation results for the proposed algorithm. The performance of the proposed algorithm is clearly degraded when compared to a simple AWGN channel. However, it performs better than all the other techniques. Interestingly, the method also also outperforms BMRC-MSDF which, as discussed in Section~\ref{subsec:complexity}, has significantly higher computational complexity.  Although this result initially appears to be counter-intuitive, further experiments where only an AWGN channel is considered or if perfect channel state information (CSI) is assumed, show comparable performance between BMRC-MSDF and EV-CSS. Therefore, we can conclude that at very low SNR the blind channel estimates based on the SVD have large errors and the full benefit of MRC is not achieved. In contrast, the EV-CSS is able to fully take advantage of the information from all antennas because the algorithm works directly with the covariance matrices instead of utilizing an estimated CSI to pre-combine the signals.

Another reason for the performance gain is that EV-CSS works directly with the time-domain cyclic autocorrelation of the PU signal which is particularly effective if some prior knowledge about the cyclic frequency $\alpha_0$ is provided. An FFT based scheme requires the use of some form of frequency smoothing which degrades the cyclic feature. However, BMRC-MSDF can be made more robust to inaccurate knowledge of the cyclic features through these smoothing techniques as described in \cite{Jitvanichphaibool2010a}. This has no bearing in the scenarios presented in this work since perfect knowledge of $\alpha_0$ is assumed and the parameters are chosen such that the cyclic frequency is perfectly aligned with an FFT bin.

The EGC approach performs worst among all the techniques in a flat-fading environment because at very low SNR the estimation of the phase of $\mathbf{h}$ has very substantial error resulting in the test statistic being degraded in most cases. In contrast to this SUM-MSDF is able to separately calculate the spectral correlation of the signal on each antenna which offers significant gain after combining.

\begin{figure}
\centering
\includegraphics[width=21pc]{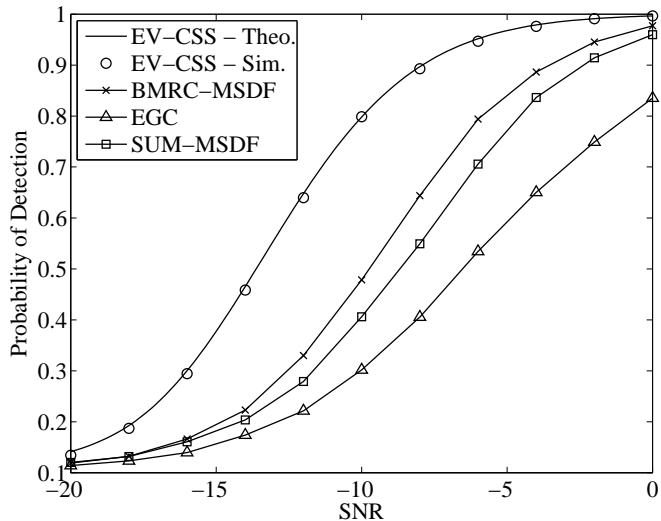}
\caption{Comparison of multiple antenna cyclostationary spectrum sensing techniques with varying SNR under Rayleigh flat-fading ($M=2$, $N=1000$, uncorrelated noise).}
\label{fig:pd_snr}
\end{figure}

The effect of varying SNR on probability of detection is also shown in Fig.~\ref{fig:pd_snr}. In both analysis and simulation plots for EV-CSS, $\gamma$ is set directly using (\ref{eqn:threshold}) in order to maintain a CFAR of $P_{\rm{FA}}=0.1$. We again observe the strong agreement between theory and simulation for EV-CSS. In addition to this, the threshold selection is shown to be very effective in achieving the desired theoretical $P_{FA}$. As for the other techniques, the threshold needs to be determined as a function of SNR. This was achieved empirically in our simulations by Monte Carlo simulations of each algorithm under $\mathcal{H}_0$. Although the distributions used for $\mathcal{H}_1$ are designed for local alternatives (very low SNR), the analysis still matches simulation very accurately. This is true even at SNR $\geq 0$ dB since the distribution converges to a non-zero mean Gaussian at very high non-centrality parameter which accurately describes the test statistic at both high $N$ and high SNR.

\subsection{Varying Sample Size and Varying Number of Antennas}

\begin{figure}
\centering
\includegraphics[width=21pc]{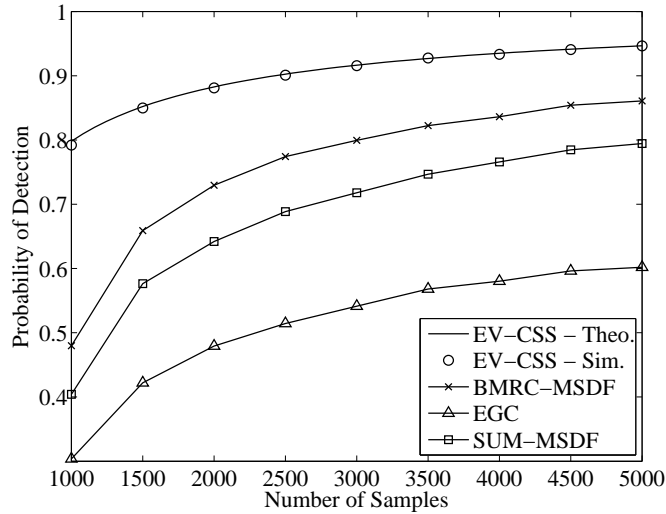}
\caption{Comparison of multiple antenna cyclostationary spectrum sensing techniques with varying sample size $N$ under Rayleigh flat-fading channel ($M=2$, $N=1000$, SNR=-10 dB, uncorrelated noise).}
\label{fig:pd_n}
\end{figure}

The probability of detection over varying number of samples $N\in[1000,5000]$ under an SNR of -10 dB and random flat-fading is shown in Fig.~\ref{fig:pd_n}. In these simulations the $P_{FA}=0.1$ and the threshold is theoretically determined for EV-CSS. For the other techniques, the CFAR threshold is determined empirically through Monte Carlo simulatoins. As with varying SNR, only one threshold calculation is done for EV-CSS due to its independence to $N$. Significant improvements in detection probability of detection can be seen with increasing sample size. However, these gains tend to slow down with higher number of samples.

\begin{figure}
\centering
\includegraphics[width=21pc]{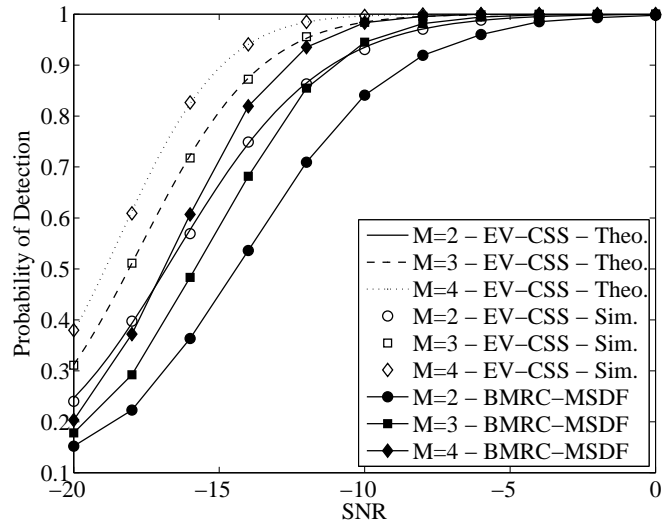}
\caption{The effect of the number of antennas on the detection probability of EV-CSS and BMRC-MSDF. The same number of samples per antenna $N=4000$ is used.}
\label{fig:varym}
\end{figure}

The effect of number of antennas, $M$, on detection accuracy is studied in Fig.~\ref{fig:varym}. In this figure only the best two algorithms (EV-CSS and BMRC-MSDF) are shown to facilitate the comparison. Note that for EV-CSS, to keep the the $P_{\rm{FA}}$ constant at 0.1 the threshold must be set to a new value based on (\ref{eqn:threshold}). On the other hand, for BMRC-MSDF, the threshold is set for different SNR and $M$. The $\sigma_\eta$ across all antennas is assumed to be the same and an SNR is set using (\ref{eqn:snr}). Since both algorithms utilize some form of blind channel estimate we expect both to perform successively better as $M$ is increased due to the improved spatial diversity provided by multiple antennas due to independent fading. Similar to previous results, the EV-CSS has better performance than BMRC-MSDF for different values of $M$ at low SNRs. The agreement between theory and simulation for EV-CSS remains very strong even with higher $M$.

\subsection{Spatially Correlated Noise}

\begin{figure}
\centering
\includegraphics[width=21pc]{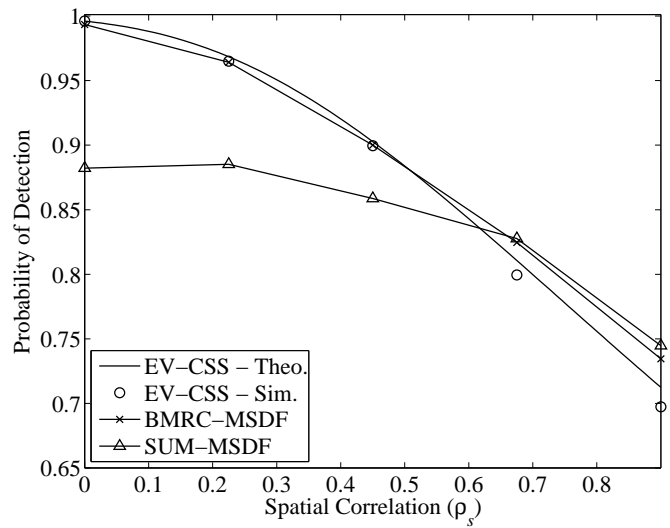}
\caption{The effect a increasing spatial correlation on detection probability. A simple AWGN channel is used in these simulations to highlight the effect of spatial correlation ($N=1000$, SNR=-10 dB).}
\label{fig:pdcorr}
\end{figure}

The effect of varying spatial correlation, $\rho_s$, is shown in Fig.~\ref{fig:pdcorr}. In this simulation we again have $M=2$ antennas. In the BMRC-MSDF simulations we assume perfect knowledge of the channel with $\mathbf{h}=[1, 1]^T$ (no fading) and therefore the EVD-based blind channel estimation is no longer performed. This also results in BMRC-MSDF and SUM-MSDF having very comparable performance and thus, only one of them is shown. All methods, are degraded by increasing levels of spatial correlation. However, we see that BMRC-MSDF is more robust to such an impairment and ends up performing slightly better than EV-CSS at very high level of correlation. However, its more likely that each antenna would have only some slight level of correlation in a properly designed RF front-end, then the two methods could be regarded as being comparable in a simple AWGN channel. SUM-MSDF is seen to be the most robust to spatially correlated noise since it calculates the spectral correlation for each antenna individually. In fact, it shows some slight improvement at $\rho_s = 0.2$.

\subsection{Robustness to Noise Uncertainty}
\begin{figure}
\centering
\includegraphics[width=21pc]{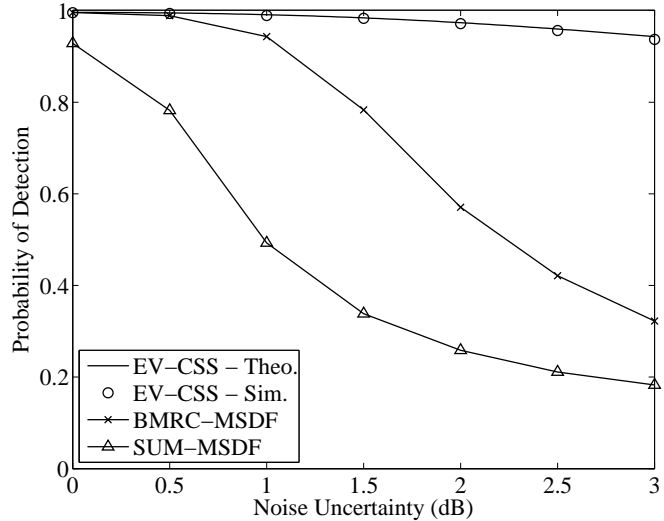}
\caption{The effect of noise uncertainty, $\Delta$, on the performance of various spectrum sensing schemes based on cyclostationarity. Noise uncertainty is assume to be uniformly distributed over an interval of $\left[\overline{\rm{SNR}}-\Delta,\overline{\rm{SNR}}+\Delta\right]$ dB. $\overline{\rm{SNR}}$=-10 dB and $N=1000$ samples (spatially uncorrelated noise)}
\label{fig:uncertainty}
\end{figure}

Finally, we test the proposed algorithm's robustness under noise uncertainty and compare it with other cyclostationary-based spectrum sensing techniques. Noise uncertatinty arises from the noise power level varying due to changes in the amount of thermal noise, amplifier gain, calibration error, and fluctuating interference \cite{Tandra2008}. The impact of the noise uncertainty is evaluated using the Bayesian statistics approach \cite{Tandra2006}, where a prior distribution on the noise power is assumed. In our simulations this is achieved by having an SNR with uniform distribution.
\begin{equation}
	f_{{\rm SNR}}\left(x\right)=\begin{cases}
\frac{1}{2\Delta}, & \overline{{\rm SNR}}-\Delta\leq x\leq\overline{{\rm SNR}}+\Delta\\
0, & {\rm otherwise}
\end{cases},
\end{equation}
where $\overline{{\rm SNR}}$ is the average SNR which we set to be -10 dB. We then calculate the $P_{\rm{FA}}$ and $P_{\rm{D}}$ by averaging over $f_{{\rm SNR}}\left(x\right)$. The results for $\Delta\in[0,3]$ dB are presented in Fig.~\ref{fig:uncertainty} where the average $P_{\rm{FA}}=0.1$. As can be seen on the figure, both BMRC-MSDF and the proposed method perform at $P_{\rm{D}}=1$ when no noise uncertainty is present. However, BMRC-MSDF and EGC performance degrades significantly even with small $\Delta$ while EV-CSS is shown to be more robust to such an impairment. The slight degradation in EV-CSS is caused by averaging $P_{\rm{D}}$ over a uniformly distributed random variable SNR where the relationship between $P_{\rm{D}}$ and SNR is non-linear. As such, the values below $\overline{{\rm SNR}}$ have a higher effect on the average $P_{\rm{D}}$ resulting in a net degradation.

%%%%%%%%%%%%%%%%%%%%%%%%%%%%%%%%%%%%%%%%%%%%%%%%%%%%%%%%%%%%%%%%%%%%%%%%%%%%%%%
%%%%%%%%%%%%%%%%%%%%%%%%%%%%%%%%%%%%%%%%%%%%%%%%%%%%%%%%%%%%%%%%%%%%%%%%%%%%%%%

\section{Conclusion}
\label{sec:Conclusion}
A multi-antenna cyclostationary-based spectrum sensing algorithm based on the cyclic correlation significance test was proposed and evaluated both analytically and through simulations. The method was shown to outperform existing multiple antenna signal-selective spectrum sensing methods in the literature. The computational complexity of the algorithm was also compared with that of the best performing existing algorithm that uses MRC by blindly estimating the CSI and was shown to require substantially less multiplications. The detection threshold for CFAR was also determined, both theoretically and via simulation, to be independent of the noise variance or the number of samples. This means that a single threshold is required for a given number of antenna, eliminating the need for separate noise estimation. The proposed method has also been shown to be highly robust to the effects of noise uncertainty.

%%%%%%%%%%%%%%%%%%%%%%%%%%%%%%%%%%%%%%%%%%%%%%%%%%%%%%%%%%%%%%%%%%%%%%%%%%%%%%%
%%%%%%%%%%%%%%%%%%%%%%%%%%%%%%%%%%%%%%%%%%%%%%%%%%%%%%%%%%%%%%%%%%%%%%%%%%%%%%%

\appendix[Derivation of True Correlation under $\mathcal{H}_1$]
\label{app:Rho}
In this appendix, we find the true correlation, $\rho$, under the non-null hypothesis $\mathcal{H}_1$ given a particular instance of the channel, $\mathbf{h}$, and noise covariance matrix, $\mathbf{R}_{\boldsymbol{\eta}\boldsymbol{\eta}}=\sigma_\eta\mathbf{A}\mathbf{A}^H$. We begin by repeating (\ref{eqn:rxx}) and (\ref{eqn:rxxalpha}) since these covariance matrices completely determine the test statistic.
\begin{eqnarray}
\label{eqn:covmat}
\mathbf{R}_{\mathbf{xx}}&=&\mathbf{h}\mathbf{h}^H+\mathbf{R}_{\boldsymbol{\eta\eta}}\\
\mathbf{R}_\mathbf{xx}^{\alpha_0}&=&\mathbf{h}\mathbf{h}^H R_{ss}^\alpha.
\end{eqnarray}
From (\ref{eqn:cclr}) and under the assumption of only one SOI ($\mu_i=0$ for $i>1$) we have
\begin{equation}
1-\mu_1^2 =  \frac{\begin{vmatrix}
      \mathbf{{R}_{xx}} & \mathbf{R}_\mathbf{xx}^{\alpha_0} \\
      \mathbf{R}_\mathbf{xx}^{\alpha_0} & \mathbf{{R}_{xx}} 
    \end{vmatrix}}{\left\vert\mathbf{{R}_{xx}}\right\vert^2},
\end{equation}
where we recognize that both covariance matrices are Hermitian symmetric. Using the determinant for block $2\times2$ block matrices, we have
\begin{eqnarray}
1-\mu_1^2 &=&  \frac{\left\vert\mathbf{{R}_{xx}}\right\vert\left\vert\mathbf{R_{xx}}-\mathbf{{R}}_\mathbf{xx}^{\alpha_0}\mathbf{{R}}_\mathbf{xx}^{-1}\mathbf{{R}}_\mathbf{xx}^{\alpha_0}\right\vert}{\left\vert\mathbf{\hat{R}_{xx}}\right\vert^2}\\
&=&\frac{\left\vert\mathbf{R_{xx}}-\mathbf{{R}}_\mathbf{xx}^{\alpha_0}\mathbf{{R}}_\mathbf{xx}^{-1}\mathbf{{R}}_\mathbf{xx}^{\alpha_0}\right\vert}{\left\vert\mathbf{{R}_{xx}}\right\vert}
\end{eqnarray}
Substituting $\mathbf{{R}}_\mathbf{xx}^{\alpha_0}$ and grouping together scalar terms
\begin{equation}
1-\mu_1^2 = \frac{\left\vert\mathbf{R_{xx}}-\left(\left\vert R_{ss}^\alpha \right\vert^2\mathbf{h}^H
\mathbf{{R}}_\mathbf{xx}^{-1}
\mathbf{h}\right)\mathbf{h}\mathbf{h}^H\right\vert}{\left\vert\mathbf{{R}_{xx}}\right\vert}
\end{equation}
Using Sylvester's determinant theorem for the sum of a full-rank and rank-1 matrix
\begin{eqnarray}
1-\mu_{1}^{2}&=&\frac{\left\vert \mathbf{R_{xx}}\right\vert \left(1-\left\vert R_{ss}^{\alpha}\right\vert ^{2}\left(\mathbf{h}^{H}\mathbf{R}_{\mathbf{xx}}^{-1}\mathbf{h}\right)^{2}\right)}{\left\vert \mathbf{R_{xx}}\right\vert}\\
\mu_1^2 &=&
\left\vert R_{ss}^\alpha\right\vert^2\left(
\mathbf{h}^H\mathbf{{R}}_\mathbf{xx}^{-1}\mathbf{h}
\right)^2\\
\mu_1 &=& \left\vert R_{ss}^\alpha\right\vert \mathbf{h}^H\mathbf{{R}}_\mathbf{xx}^{-1}\mathbf{h}
\end{eqnarray}
To proceed further, we need to evaluate the inverse of (\ref{eqn:covmat}) which is a sum of a full-rank matrix and a rank-1 matrix. Using the results in \cite[Eqn. 1]{Miller1981} the inverse becomes
\begin{equation}
\mathbf{R}_{\mathbf{xx}}^{-1} = \mathbf{R}_{\boldsymbol{\eta\eta}}^{-1} - \frac{\mathbf{R}_{\boldsymbol{\eta\eta}}^{-1}\mathbf{h}\mathbf{h}^H\mathbf{R}_{\boldsymbol{\eta\eta}}^{-1}}{1+\mathbf{h}^H\mathbf{R}_{\boldsymbol{\eta\eta}}^{-1}\mathbf{h}}.
\end{equation}
Thus we have,
\begin{eqnarray}
\mu_1 &=& \left\vert R_{ss}^\alpha\right\vert \mathbf{h}^H
\left(\mathbf{R}_{\boldsymbol{\eta\eta}}^{-1} - \frac{\mathbf{R}_{\boldsymbol{\eta\eta}}^{-1}\mathbf{h}\mathbf{h}^H\mathbf{R}_{\boldsymbol{\eta\eta}}^{-1}}{1+\mathbf{h}^H\mathbf{R}_{\boldsymbol{\eta\eta}}^{-1}\mathbf{h}}\right)
\mathbf{h}\\
&=&\left\vert R_{ss}^{\alpha}\right\vert\left( \mathbf{h}^{H}\mathbf{R}_{\boldsymbol{\eta\eta}}^{-1}\mathbf{h}-\frac{\mathbf{h}^{H}\mathbf{R}_{\boldsymbol{\eta\eta}}^{-1}\mathbf{h}\mathbf{h}^{H}\mathbf{R}_{\boldsymbol{\eta\eta}}^{-1}\mathbf{h}}{1+\mathbf{h}^{H}\mathbf{R}_{\boldsymbol{\eta\eta}}^{-1}\mathbf{h}}\right)\\
&=& \frac{\left\vert R_{ss}^{\alpha}\right\vert \mathbf{h}^{H}\mathbf{R}_{\boldsymbol{\eta\eta}}^{-1}\mathbf{h}}
{1+\mathbf{h}^{H}\mathbf{R}_{\boldsymbol{\eta\eta}}^{-1}\mathbf{h}}.
\end{eqnarray}
Recalling that $\mathbf{R}_{\boldsymbol{\eta\eta}}=\sigma_\eta \mathbf{A} \mathbf{A}^H$, we arrive at the final expression
\begin{equation}
\mu_1 = \frac{\left\vert R_{ss}^{\alpha}\right\vert \mathbf{h}^{H}\mathbf{A}^{-H}\mathbf{A}^{-1}\mathbf{h}}
{\sigma_\eta^2+\mathbf{h}^{H}\mathbf{A}^{-H}\mathbf{A}^{-1}\mathbf{h}}.
\end{equation}
%

%%%%%%%%%%%%%%%%%%%%%%%%%%%%%%%%%%%%%%%%%%%%%%%%%%%%%%%%%%%%%%%%%%%%%%%%%%%%%%%
%%%%%%%%%%%%%%%%%%%%%%%%%%%%%%%%%%%%%%%%%%%%%%%%%%%%%%%%%%%%%%%%%%%%%%%%%%%%%%%


\begin{thebibliography}{10}
\providecommand{\url}[1]{#1}
\csname url@samestyle\endcsname
\providecommand{\newblock}{\relax}
\providecommand{\bibinfo}[2]{#2}
\providecommand{\BIBentrySTDinterwordspacing}{\spaceskip=0pt\relax}
\providecommand{\BIBentryALTinterwordstretchfactor}{4}
\providecommand{\BIBentryALTinterwordspacing}{\spaceskip=\fontdimen2\font plus
\BIBentryALTinterwordstretchfactor\fontdimen3\font minus
  \fontdimen4\font\relax}
\providecommand{\BIBforeignlanguage}[2]{{%
\expandafter\ifx\csname l@#1\endcsname\relax
\typeout{** WARNING: IEEEtran.bst: No hyphenation pattern has been}%
\typeout{** loaded for the language `#1'. Using the pattern for}%
\typeout{** the default language instead.}%
\else
\language=\csname l@#1\endcsname
\fi
#2}}
\providecommand{\BIBdecl}{\relax}
\BIBdecl

\bibitem{Urriza2012}
\BIBentryALTinterwordspacing
P.~{Urriza}, E.~{Rebeiz}, and D.~{\v{C}abri\'{c}}, ``Eigenvalue-based
  cyclostationary spectrum sensing using multiple antennas,'' in \emph{Proc.
  IEEE GLOBECOM}, Anaheim, CA, USA, Dec. 3--7, 2012. [Online]. Available:
  \url{http://arxiv.org/abs/1210.8176}
\BIBentrySTDinterwordspacing

\bibitem{Haykin2005}
S.~{Haykin}, ``Cognitive radio: brain-empowered wireless communications,''
  \emph{{IEEE} J. Sel. Areas Commun.}, vol.~23, no.~2, pp. 201--220, Feb. 2005.

\bibitem{Yucek2009}
T.~{Y\"{u}cek} and H.~Arslan, ``A survey of spectrum sensing algorithms for
  cognitive radio applications,'' \emph{{IEEE} Commun. Surveys Tuts.}, vol.~11,
  no.~1, pp. 116--130, Mar. 2009.

\bibitem{Tandra2008}
R.~Tandra and A.~Sahai, ``{SNR} walls for signal detection,'' \emph{{IEEE} J.
  Sel. Topics Signal Process.}, vol.~2, no.~1, pp. 4--17, Feb. 2008.

\bibitem{Gardner1987a}
W.~A. {Gardner}, W.~A. {Brown}, and C.-K. {Chen}, ``Spectral correlation of
  modulated signals: Part {II}--digital modulation,'' \emph{{IEEE} Trans.
  Commun.}, vol.~35, no.~6, pp. 595--601, Jun. 1987.

\bibitem{Quan2008}
Z.~Quan, S.~Cui, H.~Poor, and A.~Sayed, ``Collaborative wideband sensing for
  cognitive radios,'' \emph{{IEEE} Signal Process. Mag.}, vol.~25, no.~6, pp.
  60--73, Nov. 2008.

\bibitem{Taherpour2010}
A.~{Taherpour}, M.~{Nasiri-Kenari}, and S.~{Gazor}, ``Multiple antenna spectrum
  sensing in cognitive radios,'' \emph{{IEEE} Trans. Wireless Commun.}, vol.~9,
  no.~2, pp. 814--823, Feb. 2010.

\bibitem{Tugnait2012}
J.~K. Tugnait, ``On multiple antenna spectrum sensing under noise variance
  uncertainty and flat fading,'' \emph{{IEEE} Trans. Signal Process.}, vol.~60,
  no.~4, pp. 1823 --1832, Apr. 2012.

\bibitem{Sadeghi2008}
H.~Sadeghi and P.~Azmi, ``A novel primary user detection method for
  multiple-antenna cognitive radio,'' in \emph{Proc. International Symposium on
  Telecommunications}, Aug. 2008, pp. 188 --192.

\bibitem{Chen2008}
X.~{Chen}, W.~{Xu}, Z.~{He}, and X.~{Tao}, ``Spectral correlation-based
  multi-antenna spectrum sensing technique,'' in \emph{Proc. IEEE WCNC}, Las
  Vegas, NV, USA, Mar. 31--Apr. 3, 2008.

\bibitem{Jitvanichphaibool2010a}
K.~{Jitvanichphaibool}, Y.-C. {Liang}, and Y.~{Zeng}, ``Spectrum sensing using
  multiple antennas for spatially and temporally correlated noise
  environments,'' in \emph{Proc. IEEE DySPAN}, Singapore, Apr. 6--9, 2010.

\bibitem{Schell1990a}
S.~{Schell} and W.~{Gardner}, ``Detection of the number of cyclostationary
  signals in unknown interference and noise,'' in \emph{Proc. ACSSC}, Pacific
  Grove, CA, USA, Nov. 5--7, 1990.

\bibitem{Shrimpton1997}
T.~Shrimpton and S.~Schell, ``Source enumeration using a signal-selective
  information theoretic criterion,'' in \emph{Proc. MILCOM}, Nov. 2--5, 1997.

\bibitem{Lawley1959}
D.~N. {Lawley}, ``Tests of significance in canonical analysis,''
  \emph{Biometrika}, vol.~46, no. 1/2, pp. 59--66, Jun. 1959.

\bibitem{Anderson2003}
T.~W. Anderson, \emph{An Introduction to Multivariate Statistical Analysis},
  3rd, Ed.\hskip 1em plus 0.5em minus 0.4em\relax Wiley, 2003.

\bibitem{Bartlett1938}
M.~S. Bartlett, ``Further aspects of the theory of multiple regression,''
  \emph{Mathematical Proceedings of the Cambridge Philosophical Society},
  vol.~34, no.~01, pp. 33--40, 1938.

\bibitem{Sugiura1969}
N.~Sugiura and Y.~Fujikoshi, ``\BIBforeignlanguage{English}{Asymptotic
  expansions of the non-null distributions of the likelihood ratio criteria for
  multivariate linear hypothesis and independence},''
  \emph{\BIBforeignlanguage{English}{The Annals of Mathematical Statistics}},
  vol.~40, no.~3, pp. 942--952, 1969.

\bibitem{Bartlett1947}
M.~S. Bartlett, ``\BIBforeignlanguage{English}{The general canonical
  correlation distribution},'' \emph{\BIBforeignlanguage{English}{The Annals of
  Mathematical Statistics}}, vol.~18, no.~1, pp. 1--17, 1947.

\bibitem{Sugiura1973}
N.~Sugiura, ``\BIBforeignlanguage{English}{Asymptotic non-null distributions of
  the likelihood ratio criteria for covariance matrix under local
  alternatives},'' \emph{\BIBforeignlanguage{English}{The Annals of
  Statistics}}, vol.~1, no.~4, pp. 718--728, 1973.

\bibitem{Muirhead1980}
R.~J. Muirhead and C.~M. Waternaux, ``Asymptotic distributions in canonical
  correlation analysis and other multivariate procedures for nonnormal
  populations,'' \emph{Biometrika}, vol.~67, no.~1, pp. 31--43, 1980.

\bibitem{Ogasawara2007}
H.~Ogasawara, ``Asymptotic expansions of the distributions of estimators in
  canonical correlation analysis under nonnormality,'' \emph{Journal of
  Multivariate Analysis}, vol.~98, no.~9, pp. 1726--1750, 2007.

\bibitem{Rebeiz2011}
E.~{Rebeiz} and D.~{\v{C}abri\'{c}}, ``Low complexity feature-based modulation
  classifier and its non-asymptotic analysis,'' in \emph{Proc. IEEE GLOBECOM},
  Houston, TX, USA, Dec. 5--9, 2011.

\bibitem{Zeng2009}
Y.~{Zeng} and Y.-C. {Liang}, ``Eigenvalue-based spectrum sensing algorithms for
  cognitive radio,'' \emph{{IEEE} Trans. Commun.}, vol.~57, no.~6, pp.
  1784--793, Jun. 2009.

\bibitem{Tandra2006}
R.~Tandra and A.~Sahai, \emph{Performance of the Power Detector with Noise
  Uncertainty}, IEEE 802.22-06/0075r0 Std., Jul. 2006.

\bibitem{Miller1981}
K.~S. Miller, ``\BIBforeignlanguage{English}{On the inverse of the sum of
  matrices},'' \emph{\BIBforeignlanguage{English}{Mathematics Magazine}},
  vol.~54, no.~2, pp. 67--72, 1981.

\end{thebibliography}
\end{document}